\documentclass[aps,pra,twocolumn,superscriptaddress,showpacs]{revtex4}
\usepackage{bm}
 
\usepackage{amssymb} 

\usepackage{graphicx}

\begin{document}

\title{A planar multipole ion trap}

\author{M. Debatin}
\affiliation{Physikalisches Institut, Universit{\"a}t Freiburg,
Hermann-Herder-Str.\ 3, 79104 Freiburg, Germany}
\author{M. Kr{\"o}ner}
\affiliation{Department of Microsystem Engineering (IMTEK), Universit{\"a}t 
Freiburg, Georges-K{\"o}hler-Allee 102, 79110 Freiburg, Germany}
\author{J. Mikosch}
\author{S. Trippel}
\author{N. Morrison}
\altaffiliation{Permanent address: Dept. of Physics, Carnegie Mellon University, Pittsburgh, PA 15213, USA}
\author{M. Reetz-Lamour}
\affiliation{Physikalisches Institut, Universit{\"a}t Freiburg,
Hermann-Herder-Str.\ 3, 79104 Freiburg, Germany}
\author{P. Woias}
\affiliation{Department of Microsystem Engineering (IMTEK), Universit{\"a}t 
Freiburg, Georges-K{\"o}hler-Allee 102, 79110 Freiburg, Germany}
\author{R. Wester}
\email{roland.wester@physik.uni-freiburg.de}
\author{M. Weidem{\"u}ller}
\email{m.weidemueller@physik.uni-freiburg.de}
\affiliation{Physikalisches Institut, Universit{\"a}t Freiburg,
Hermann-Herder-Str.\ 3, 79104 Freiburg, Germany}

\date{\today}


\begin{abstract}
We report on the realisation of a chip-based multipole ion trap manufactured
using micro-electromechanical systems (MEMS) technology. It provides ion
confinement in an almost field-free volume between two planes of
radiofrequency electrodes, deposited on glass substrates, which allows for
optical access to the trap. An analytical model of the effective trapping
potential is presented and compared with numerical calculations. Stable
trapping of argon ions is achieved and a lifetime of 16\,s is
measured. Electrostatic charging of the chip surfaces is studied and found to
agree with a numerical estimate.
\end{abstract}


\maketitle


\section{introduction}

Microchip-based ion traps are being investigated in several laboratories
worldwide for purposes ranging from mass spectrometry
\cite{blain2004:ijm,shortt2005:jms} to quantum information
\cite{seidelin2006:prl,pearson2006:pra,stick2006:nat}. Such traps can be
precisely manufactured using micro-electromechanical systems (MEMS) technology
offering highly integrated setups. Radiofrequency paul traps are being
developed with ions trapped above the surface of a single chip
\cite{pearson2006:pra,seidelin2006:prl} or between electrodes placed on
different chips \cite{schulz2006:fp,brownnutt06}. For loading these traps,
photoionisation techniques using laser-ablated gas \cite{hendricks2007:apb} or
laser-cooled neutral atoms \cite{cetina2007:arxiv} are utilized.

Here we present a planar microchip-based ion trap with a multipole arrangement
of radiofrequency electrodes. Built from classically machined components, such
multipole ion traps, in particular the 22-pole trap \cite{gerlich1995:ps}, are
successfully used for the study of low-temperature ion-molecule reactions of
astrophysical interest \cite{paul95,gerlich2006:ps} and to investigate
laser-induced reaction processes \cite{schlemmer2002:jcp,asvany2005:sci,
mikosch2004:jcp,trippel2006:prl,dzhonson07}. The multipole structure leads to
an effective potential with a finite depth and a large field-free central
region \cite{gerlich1995:ps,trippel2006:prl,mikosch2007:prl} that allows for
buffer gas thermalization of the translational and rovibrational degrees of
freedom of trapped molecular ions \cite{glosik06,mercier06,mikosch2004:jcp}.
We have transformed the cylindrical design of a conventional 22-pole trap into
a planar electrode structure, which allows for MEMS fabrication. The open
geometry of this planar configuration, and the application of transparent
indium tin oxide (ITO) electrodes, will allow us to overlap an optically
trapped cloud of ultracold atoms with ions confined in the microchip-based
trap. This will open up opportunities for sympathetic cooling of ions with
ultracold atoms and for experimental investigations of ultracold ion-atom
interactions.

In this work, the operation of the planar trap and its characteristics are
described. Numerical simulations of the trapping field and details of the MEMS
process will be described elsewhere \cite{kroener:s_a}. The paper is organized
as follows: an analytical model of the effective potential of the chip-based
multipole trap is presented in the next section, followed by a description of
the trap setup in section \ref{setup:sect}. Experimental results on ion
trapping and on the achieved trap lifetimes are discussed in section
\ref{results:sect}. The analysis of surface charging effects are presented in
section \ref{charging:sect}.


\section{\label{design:sect} Properties of the chip-based multipole ion trap}

\begin{figure}
\includegraphics[width=\columnwidth]{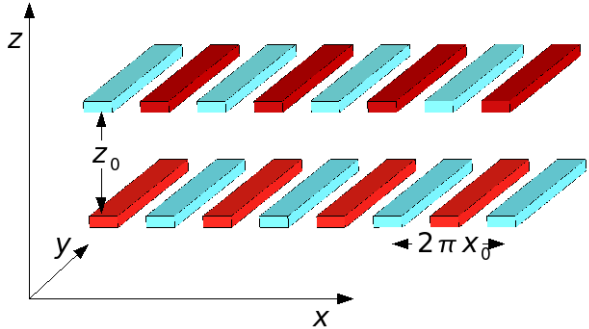}
\caption{\label{falleschema} (Color online) Schematic view of the planar
multipole ion trap with equidistant electrodes in two nearby planes. The
electrodes are alternatingly connected to two opposing radiofrequency
potentials to provide confinement between the planes.}
\end{figure}

The basic components of the planar chip-based multipole ion trap are two sets
of equally spaced and equally broad conducting stripes deposited on two
insulating glass substrates that face each other. Fig.\ \ref{falleschema}
shows a schematic view of the trap; every second stripe is connected to an
rf-potential $U_0 \sin(\omega t)$ and the other stripes are connected to the
opposing rf-potential $-U_0 \sin(\omega t)$. As shown below, this leads to a
repulsive effective potential in front of each of the two electrode planes,
thus yielding confinement of ions between the two planes. The distance from
the center of one stripe to the center of the next one is given by $\pi x_0$
and the distance between the two substrate surfaces is denoted $z_0$. The
width of the stripes is assumed to be $\pi x_0 / 2$. In our realisation $\pi
x_0 = 1$\,mm and $z_0 = 5$\,mm is employed.

For an analytical description of the potential generated by the two planes of
radiofrequency electrodes we assume the plane to carry an infinite number of
stripes and the stripes to extend infinitely in the plane. We further assume
quasistationary conditions, a good approximation for trap frequencies in the
MHz regime, and obtain the potential $\Phi(\vec r) \sin(\omega t)$ by solving
the Laplace equation
\begin{equation}
\Delta \Phi(\vec r)=0.
\end{equation}
Fig.\ \ref{falleschema} shows the employed coordinate system. The boundary
conditions of the periodic arrangement of stripes are given by a periodic
trapezoidal function: the potential is constant along the electrode surfaces
and linear between the electrodes. This potential is approximated by the first
order term of its Fourier series which reads $U(x, z=\pm z_0/2, t) = 1.15\,U_0
\cos(x/x_0) \sin(\omega t)$. This approximate boundary condition satisfies the
requirement of opposite voltages on neighbouring electrodes. For distances
$\Delta z > x_0$ from the trap electrodes it is a good approximation, as shown
below. For these boundary conditions an analytical solution for the electric
field inside the trap is given by,
\begin{equation}
\Phi(\vec r)=\Phi_0 \sinh(\hat{z}) \cos(\hat{x}),
\end{equation}
where $\hat{z}=z/x_0$ and $\hat{x}=x/x_0$ are reduced variables. The value of
$\Phi_0$ is linked to the potential $U_0$ applied to the electrodes by $\Phi_0
= 1.15\,U_0 / \sinh[z_0/(2 x_0)]$.

The effective potential that an adiabatically trapped ion experiences in a
rapidly oscillating rf field is given by \cite{gerlich1992:adv}
\begin{equation}
V^*(\vec r)=\frac{q^2}{4 m \omega^2} [\vec \nabla \Phi(\vec r)]^2,
\label{effective_potential}
\end{equation}
where the charge and mass of the ion are denoted as $q$ and $m$. For the given
solution for the chip-based ion trap this yields
\begin{equation}
V^*(\vec r)=
\frac{(1.15)^2 q^2 U_0^2}{4 m \omega^2 x_0^2}
\frac{\cosh(2 \hat{z}) + \cos(2 \hat{x})}
     {\cosh(z_0/x_0)-1}.
\end{equation}
For $z \gg x_0$ this solution is approximately proportional to $\exp(2
\hat{z})$. This is in contrast to cylindrical multipole ion traps of order
$n$, such as the 22-pole trap ($n=11$) \cite{gerlich1995:ps}, which feature
effective potentials proportional to $r^{(2n-2)}$.

\begin{figure}
\includegraphics[width=\columnwidth]{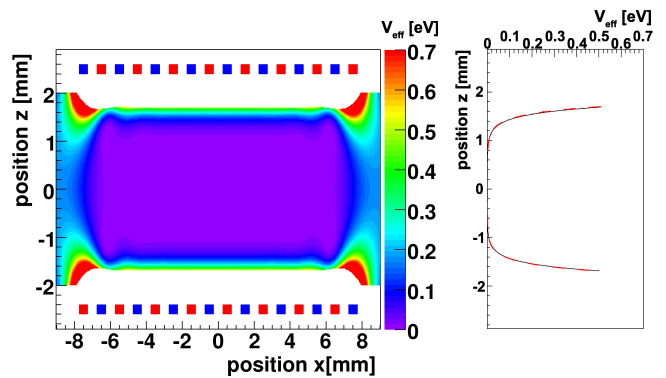}
\caption{\label{pottopf} (Color online) The right panel shows the analytically
calculated effective trapping potential along the z-direction (black
line). The one-dimensional cut obtained from a two-dimensional numerical
calculation of the effective potential (red line) can not be distinguished
from the analytical model. The result of the two-dimensional calculation is
shown as a contour plot in the left panel for 16 stripes on each plane. One
can clearly see the flat bottom and the steep walls of the effective
potential. The nonadiabatic regions, where no stable trapping is possible, are
colored in white.}
\end{figure}

The necessary condition of adiabatic motion for a trapped ion in a
time-varying field is characterized by the adiabaticity parameter
\cite{gerlich1992:adv}
\begin{equation}
\eta(\vec r)=\frac{2q 
\left |
\vec \nabla |\vec \nabla \Phi(\vec r) | 
\right |
} {m \omega^2},
\end{equation}
Ref.\ \cite{gerlich1992:adv} postulates that $\eta$ has to be less than 0.3 to
guarantee ``safe operating conditions''. We have thoroughly investigated trap
loss out of multipole traps \cite{mikosch2007:prl} and found trapping to occur
up to a value of 0.38 for $\eta$. Where $\eta$ reaches this maximum value the
surface of the trapping volume is reached. The effective potential on this
surface represents the maximum potential depth for trapped ions
\cite{mikosch2007:prl}. The right panel of Fig.\ \ref{pottopf} shows the
effective trapping potential of the chip-based multipole ion trap in the
region of space where adiabatic trapping is possible, i.\ e.\ where the
adiabaticity criterion of $\eta < 0.38$ is fulfilled. The potential is
calculated for Ar$^+$ ions in a trap of amplitude $U_0 = 125$\,V and frequency
$\omega = 2 \pi \times 5.75$\,MHz. It can be seen that the effective potential
is represented by a deep well with an almost flat, field-free bottom and with
exponentially rising potential walls and a height of about 0.5\,eV.

The electric field configuration for stable ion trapping has also been
investigated in numerical simulations and the resulting effective trapping
potentials and $\eta$-parameters are evaluated \cite{kroener:s_a}. From a
two-dimensional simulation of the effective trapping potential using SIMION
\cite{simion7}, a one-dimensional cut along the $z$-direction in the center of
the trap (for $x=y=0$) is derived. It cannot be distinguished from the
analytical model in the right panel of Fig.\ \ref{pottopf}. Both results are
found to agree within one percent, which proves the applicability of the
analytical model in the region of the trap where adiabatic motion
prevails. The full two-dimensional calculation in the $xz$-plane is shown in
the left panel of Fig.\ \ref{pottopf}. We find that the confinement in the
$z$-direction is independent of the $x$-position for almost the entire
trap. One can also see that the confinement for small and large $x$-values is
not provided by the rf fields. The same holds for small and large
$y$-values. Confinement in the $xy$-plane is therefore achieved by
superimposing additional electrostatic potentials.


\section{\label{setup:sect} Realisation of the trap and loading scheme}

\begin{figure}
\includegraphics[width=\columnwidth]{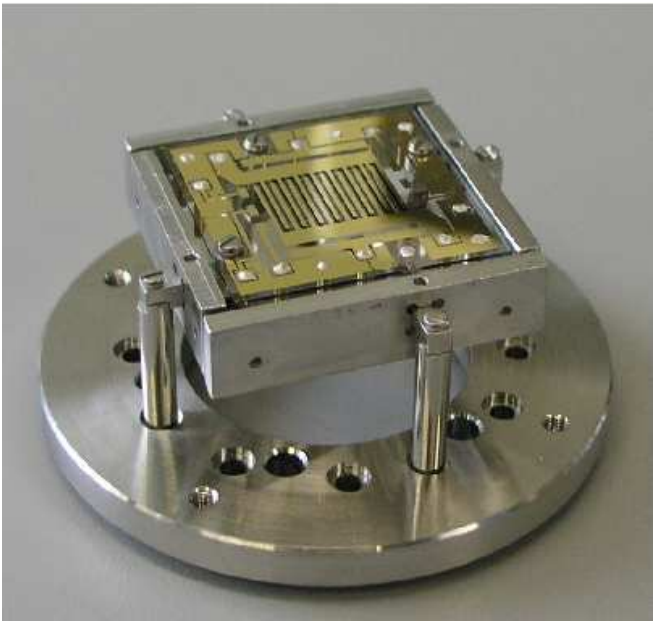}
\caption{\label{fallefoto} Photograph of one of the two ion trap chips mounted
into its holder. The second chip (not shown) is mounted 5\,mm above, facing
the first chip. The metal bars surrounding the chip serve to shield the chip
from electrostatic charging.}
\end{figure}

Two planes of gold electrodes on top of two glass substrates that face each
other form the ion trap. Design and fabrication of the chip-based ion trap
using MEMS technology will be described in a separate publication
\cite{kroener:s_a}. Fig.\ \ref{fallefoto} shows a picture of one of the two
glass substrates with the rf electrodes, spaced at $\pi x_0 = 1$\,mm, and
several static electrodes surrounding the comb structure. Besides providing
three-dimensional trapping these static electrodes are also used for the
controlled extraction of trapped ions. The second glass chip is mounted facing
the first one at a distance of $z_0 = 5$\,mm.

The trap is kept in a vacuum chamber at a residual gas pressure of about
$10^{-8}\mathrm{mbar}$ generated by a $500\,\mathrm{l/s}$ turbo molecular
pump. It is mounted in a holder fixed at one flange which also supports the
electrical connections for the trap. The radiofrequency amplitude of the trap
is generated by amplifying the signal of an rf oscillator (Hameg HM8032) in a
high frequency power amplifier (RFPA RF001100-8). To reach sufficiently high
amplitudes the output is transformed by a coil on a high frequency ferrite
core located close to the trap outside the chamber. In this way peak
amplitudes of $U_0=0...250$\,V and frequencies in the range of $\omega/(2\pi)
= 3 ... 6.5$\,MHz are applied.

Ions are created by electron impact on neutral atoms inside the trap. This is
achieved by crossing a pulsed gas beam from a piezoelectric valve
\cite{gerlich_PV} with a pulsed 1\,kV electron beam in the center of the
trap. Creating the ions inside the trap is favored over ion transport and
capturing techniques due to its simplicity but causes charging of non
conducting parts (see section \ref{charging:sect}) as well as a higher
background pressure for the first tenths of ms after the pulse. When ions are
created the electron beam is adjusted by optimizing the ion signal on a
channeltron detector, which is mounted opposite of the pulsed valve and is set
up to detect and amplify individual ion pulses.  The number of ions hitting
the detector are measured using a single channel discriminator and a
counter. Large numbers of trapped ions are measured by digitizing the current
signal of the channeltron with an oscilloscope. The data acquisition timing is
controlled with an AVR Atmel microprocessor (AT90S8515).


\section{\label{results:sect} Characterization of the trap}

Operation of the planar ion trap with Ar$^+$ ions has been achieved with the
design parameters for the rf and dc potentials obtained from the numerical
simulations, i.\ e.\ $\omega=2\pi\times 5.75$\,MHz and $U_0=125$\,V. The best
operating conditions are found by optimizing the electrostatic electrodes
surrounding the trap. These optimal settings result in static voltages of up
to a few volts. The setup is found to be stable against slight variations of
single static potentials: varying the static potentials by less than 1 Volt
from their optimum values decreases the lifetime due to a lower potential
depth, but trapping is still possible.

For extraction the potential of the surrounding border electrode in the
direction of the detector is lowered to -15\,V. More negative extraction
potentials lead to a decrease in ion signal as the ions are hitting the
electrode. More positive extraction potentials lead to a smearing of the ion
signal in the time domain as the ions close to the border are accelerated by
the extraction potential but the ions further away are much less influenced.
In experiments with few trapped ions up to 200 individual ions are counted,
limited by the overlapping ion signals in the counter. We use these data to
calibrate the analog current signal of the channeltron detector to the ion
number. In this way the largest observed analog signals of trapped ions are
found to contain about 3000 ions.

\begin{figure}
\includegraphics[width=\columnwidth]{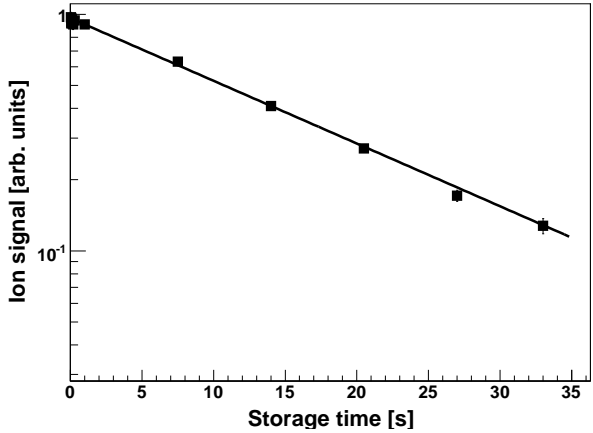}
\caption{\label{speicherkurve} Number of ions extracted from the trap after
different storage times. The solid line is showing an exponential fit with a
lifetime of 16\,s.}
\end{figure}

For the ion trap we determine a storage time of 16\,s, which corresponds to a
loss rate of 0.06\,s$^{-1}$. This lifetime can be compared to the
evaporation limited lifetime over the rim of the trapping potential
\cite{mikosch2007:prl}: The evaporation rate is given by
\begin{equation}
k(T) = A e^{-E_a / k_B T},
\end{equation}
where the trap depth $E_a \approx 0.5$\,V is taken from the effective
potential calculation of section \ref{design:sect}. The temperature of the
trapped ions is estimated to be roughly room temperature, controlled by
collisions of the trapped ions with the gas injected into the trap chamber
after ion formation. The pre-factor $A$ is assumed to be similar to the value
obtained in the 22-pole ion trap, $A=10^7$\,s$^{-1}$
\cite{mikosch2007:prl}. This yields a value of about 0.02\,s$^{-1}$, which is
only a factor of three away from the measured storage time. This is considered
a fair agreement when keeping in mind the exponential dependence of the
evaporation rate on the trap depth $E_a$.


\section{\label{charging:sect} Electrostatic charging}

Avoiding stray charges and investigating their effects where they can not be
completely eliminated is a central issue in the design of micro trap
structures where conducting and nonconducting areas are lying close to each
other and to the center of the trap \cite{seidelin2006:prl}. In our current
trap design these charging effects are non-negligible and affect both trapping
efficiency and storage time.

\begin{figure}
\includegraphics[width=\columnwidth]{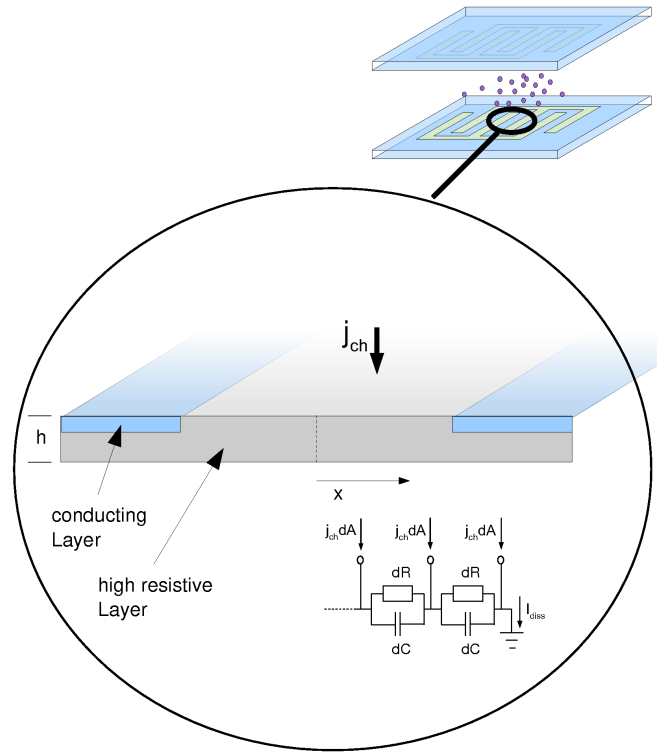}
\caption{\label{auflad} The picture is showing a cut through the chip. A high
resistive region (substrate) is enclosed by two conducting parts
(electrodes).}
\end{figure}

The steady-state potential of the glass surface induced by charging can be
calculated assuming a constant current density $j_{\rm ch}$ that is flowing
onto the surface and a resistivity-limited discharging current $I_{\rm dis}$
within the glass (see Fig.\ \ref{auflad}). The current inside the surface
flows from the middle of the high-resistive region (denoted as $x=0$) to the
two neighbouring electrode stripes. A surface area $x\,\Delta y$ (with $0 < x
< \pi x_0 / 4$ perpendicular to the stripes and $\Delta y$ parallel to the
stripes) leads to a discharging current at the position $x$ inside the glass
of
\begin{equation}
I_{\rm dis}(x)=j_{\rm ch} x\,\Delta y.
\end{equation}  
Under steady state conditions only the resistivity of the glass substrate and
not the parallel capacity determines the potential (see equivalent circuit
diagram in Fig.\ \ref{auflad}). This leads to a potential gradient at a
position $x$ between the stripes of
\begin{equation}
\frac{dU}{dx}=\frac{\rho}{h\,\Delta y} I_{\rm dis}(x)
\end{equation} 
where the discharching current $I_{\rm dis}$ flows through the area $h\,\Delta
y$ in the glass chip and $\rho$ denotes the specific resistance of the glass.
Integration from $x=0$ to $x=\pi x_0 / 4$ yields the electric potential at the
center of the high-resistive region ($x=0$) of
\begin{equation}
U_{\rm ch}=\frac{1}{2} \frac{\rho}{h} j_{\rm ch} \frac{(\pi x_0)^2}{16},
\end{equation}
with respect to the electrodes. For an estimation of the amount of charge
needed to significantly influence storage of ions we assume that a potential
of 500\,mV between two rf electrodes, a value similar to the depth of the
effective potential, will preclude trapping of ions. The resistivity $\rho$ of
the glass substrate (thickness $h=0.05$\,cm) is extrapolated from the material
data sheet \cite{borofloat} to $\approx 10^{15}\mathrm{\Omega cm}$. Thus, a
potential of 500\,mV is obtained for a charging current density of about
$5\times 10^5$ electrons per cm$^2$ and second. For the two entire chips with
their total glass surface of $2 \times 4.5$\,cm$^2$, this means that a charge
flux of about $5\times 10^6$ elementary charges per second will have a
significant influence on trapping and storage. At the typical repetition rate
of the experiment of 10 cycles per second, this yields a maximum allowable
current of $5\times 10^5$ charges per trap loading.

\begin{figure}
\includegraphics[width=\columnwidth]{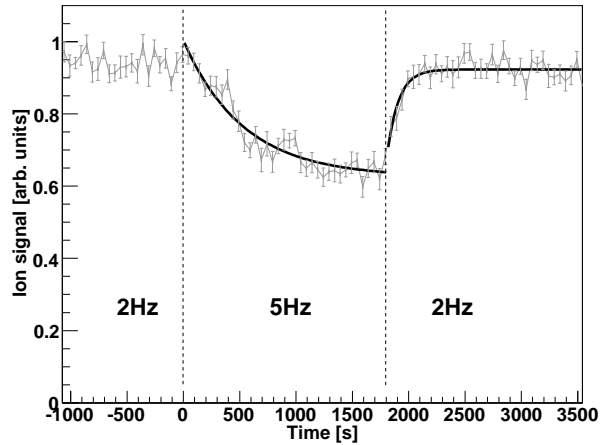}
\caption{\label{auflad2} The picture is showing a decrease in trapping
efficiency after changing from 2 loading cycles per second to 5 loading cycles
per second. The increase on the right side of the graph is the result of
switching back again to two loading cycles per second.}
\end{figure}

To investigate charging effects of the planar ion trap experimentally, the
trapping efficiency is measured for different average currents of the electron
beam used for ionization. We define the trapping efficieny as the number of
ions trapped after 10\,ms of storage time. This time is much shorter than the
lifetime of trapped ions but is also long enough to allow for complete
randomization of ion trajectories. In the experiment, charging of the chips'
surface stems from the electron beam, which is pulsed on only during loading
of the trap. The average charging current can therefore be varied by changing
the repetition rate of trap loading from 0.1\,Hz to 20\,Hz. The trapping
efficiency is measured for many trapping cycles over a time span of several
hours. In Fig.\ \ref{auflad2} the change in the trapping efficiency is shown
when the repetition rate is changed from 2\,Hz to 5\,Hz and back to
2\,Hz. With the higher repetition rate the charging increases and consequently
the trapping efficiency decreases until the repetition rate is set back to
2\,Hz and the charging is reduced again. The time constants for reaching
steady-state trapping efficiencies upon increased and decreased surface
charging are obtained by fitting a decay curve $A\exp(-t/\tau)+B$ and a growth
curve $A(1-\exp(-t/\tau'))+B$ to the data (solid line in Fig.\ \ref{auflad2}).
The obtained values for increased and decreased charging amount to $\tau
\approx 600$\,s and $\tau' \approx 100$\,s, respectively. The observation of
two different values may indicate that the increased charging is limited by
the current $j_{\rm ch}$ whereas the decreased charging is only limited by the
intrinsic capacitance and resitivity of the substrate.

To estimate the expected time constant for discharging the surface we use the
equivalent circuit of the chip surface shown in Fig.\ \ref{auflad}. The time
constant $\tau = dR\,dC$ that determines changes of the steady-state potential
depends on the resistance $dR=\frac{\rho}{h\,\Delta y} dx$ and the capacity
$dC=\epsilon_0 \epsilon_r\frac{h\,\Delta y}{dx}$. This yields the time constant
\begin{equation}
\tau=\epsilon_0 \epsilon_r \rho.
\end{equation} 
With $\epsilon_r=4.6$ for the glass substrate \cite{borofloat} one obtains a
typical time constant of about 400\,s for changes of the charging potential of
the glass substrate. Under the assumption that small changes of the trapping
efficiency are to a first approximation proportional to small changes in the
charging potential one can compare this calculated time constant to the values
obtained from the measured trapping efficiency. The order of magnitude
agreement that one finds provides evidence that charging of the glass surface
is in fact the major cause for the observed changes in the trapping
efficiency. Decreasing the resistivity of the glass substrate by an order of
magnitude one can proportionally reduce the charging potentials of the
substrate to an insignificant amount, while still maintaining small resistive
losses for the driving rf amplitude.


\section{Conclusions and outlook}

We have presented a chip-based multipole ion trap based on a planar design,
which features a large field free trapping volume between two glass substrates
carrying stripes of radiofrequency electrodes. An analytical model has been
presented that describes the effective trapping potential in good agreement
with numerical calculations. Trapping of ions has been demonstrated and the
measured decay rate of trapped Ar$^+$ ions follows the expectations from
evaporative losses over the rim of the confining potential. The effect of
surface charging, due to the highly resistive glass substrates, on the ion
trapping efficiency has been experimentally studied. The charging potential
and the observed time constant for reaching steady-state conditions has been
successfully modeled using an appropriate equivalent circuit, which is based
on the resistivity and capacity of the glass substrate.

As a next step we will add a drift tube for the extracted ions to implement a
Wiley McLaren \cite{WileyMcLaren} type time of flight mass spectrometer. To
characterize the density distribution of the trapped ions, photodetachment
tomography experiments \cite{trippel2006:prl} will be carried out. Further
improvements of the design and the fabrication techniques of the trap are
under development, including electrode materials with high optical
transmission \cite{kroener:s_a}. This will allow the combination of the
chip-based multipole ion trap with a magneto optical trap for ultracold
neutral atoms for experiments on interactions of trapped ions and clusters
with ultracold atoms.


\begin{acknowledgments}
This project is supported in part by a grant from the Ministry of Science,
Research and Arts of Baden-W{\"u}rttemberg. The chips were fabricated in the
Clean Room Service Center of the Department of Microsystems Engineering
(IMTEK), Freiburg. N. M. acknowledges support from the RISE program of the
German Academic Exchange Service (DAAD).
\end{acknowledgments}



\end{document}